\documentstyle[preprint,eqsecnum,aps]{revtex}
\begin{document}
\draft
\preprint{HYUPT-94/05 SNUTP 94-84\hspace{-26.5mm}
\raisebox{2.4ex}{cond-mat/9409025}}
\title{Generalized Laughlin Wave Functions \\
Including the Effect of Coulomb Interaction}
\author{O. J. Kwon}
\address{
Department of Physics, Pai-Chai
University, Taejon 302-162, Korea \\
{\rm and}\\
Department of Physics, Hanyang University, Seoul 133-791, Korea}
\author{Bum-Hoon Lee and Sang-Jin Sin}
\address{
Department of Physics, Hanyang University, Seoul 133-791, Korea
}
\maketitle
\begin{abstract}
We present improved wave functions for the ground state, Laughlin
quasihole and quasiparticle excitations of the fractional quantum Hall effect.
These depend explicitly on the effective strength of Coulomb interaction
and reproduce Laughlin's original result in the limit of no
Coulomb interaction.
\end{abstract}

\narrowtext
The fractional quantum Hall effect (FQHE) has attracted much interest since
it was discovered in the 2D system of electrons subject to a high
perpendicular magnetic field \cite{Tsui82}. Laughlin first
proposed a liquid-type ground state wave function (GSWF) \cite{Laug83}.
His wave functions for the ground state, quasihole and quasiparticle
excitations  are independent of the
effective strength of the inter-electron
Coulomb interaction. However it has been
observed in experiments that the energy gap
is greatly reduced as the effective strength of Coulomb interaction increases
\cite{Sant92}.

After Laughlin's
seminal work, most of the discussions on the fractional
quantum Hall effect
using a microscopic trial wave function has been confined
to the strong magnetic field limit, at which the effect of
Landau level mixing can be ignored.
The only exception is the recent work by Price et. al. \cite{Pric93}
on the spherical geometry.
Their trial wave function includes a term  which is similar to
 the pseudopotential proposed
by Ceperley \cite{Cepe78} to include the effect of
Coulomb interaction on the
Wigner crystalization of the fermion one-component plasma.

We present in this Letter a ground state wave function depending explictly on
the Coulomb interaction, which is derived in a plausible manner from
a Chern-Simons
gauge field theory in the plane geometry. The modifying term is different
from the term in Ref. \cite{Cepe78}.
Once a ground state wave function is given and if it
is nondegenerate, wave functions for Laughlin quasihole and quasiparticle
excitations can be written down directly following the Laughlin's argument of
adiabatic flux insertion. To support our reasoning, we also have done
numerical calculations. We have compared numerically the Laughlin ground
state wave function and our trial wave function in the case of up to five
particles. This
calculation shows the superiority of our trial wave function. This positive
result provides a motivation for further numerical efforts.

Let's start  with   a  non-relativistic Chern-Simons gauge
field  theory.
The Hamiltonian is
\begin{equation}
H =\int   d^2   x  \left(
 \frac{1}{2m}   |D_j   \psi   |^2  +  \frac{\alpha}{m}
  ( \psi^{\dagger} \psi )^2  + V_c ( \psi ) \right),
\end{equation}
where $j=1,2,$ $m$  is the effective mass of   charge  carrier (electron
or hole), $i D_j = i \partial_j +  a_j  -  q A_j$ and
$V_c ( \psi )$ is the Coulomb interaction term given by
\begin{equation}
V_c(\psi) = \frac{q^{2}}{2 \epsilon} \int  d^2  y   \,
(   \psi^{\dagger}  \psi (x) - {  \rho} )
 \frac{1}{|x-y|} ( \psi^{\dagger} \psi (y) - {  \rho} ),
\end{equation}
where  $q$ is  the charge    of  the particle,
$\epsilon$  is the dielectric  constant, ${  \rho}$ is  the
particle density of   uniform neutralizing   background  charges, and
 $\psi^\dagger\psi $
 is  the particle density of charge carriers whose mean value is $\rho$.
We
use  the    unit such that $c=\hbar =1$.
In two spatial dimensions a  spin polarized fermion
can be traded with a  composite  of   a    hard-core
boson and  odd   integer   number   of  flux quanta \cite{Sen91}.
We realize hard-core bosons by
bosons interacting each other with a  repulsive $\delta$-function   type
interaction,  which gives  rise to the term
$   (   \psi^{\dagger}  \psi    )^2  $.
The interaction between the attached fluxes
can be described  by the Chern-Simons    gauge     field  $a_j$
 satisfying  the Gauss constraint equation
 $(1/2   \alpha  ) \epsilon_{ij}  \partial_i    a_j = \psi^{\dagger} \psi$
 where  $\epsilon_{12} =-\epsilon_{21}     =1$. Here $\alpha$ is the odd
integer $n$ times $\pi$, which is called the statistics parameter
\cite{Lein77}.
If  $B$ is the strength  of   external  magnetic  field,
the gauge potential $A_j=-(B/2) \epsilon_{jk} x^{k}$ in the symmetric   gauge.
Terms in the Hamiltonian are
understood to be normally ordered. We will not specify the operator orderings
here since the effects to the ground state wave function are irrelevant
for our purpose.
We emphasize that our model is not the one conventionally
used in the effective field theory of the fractional quantum Hall effect
\cite{Girv90}.
In the effective theory the hard-core interaction term comes from
the Coulomb interaction in the microscopic theory and thus there is
no explicit Coulomb interaction term in the effective theory \cite{Zhan89}.
However in our derivation the hard-core
interaction term turns out to have no relevance to the ground
state wave function as shown by Haldane \cite{Hald83}. Hence
we will adopt the second quantized version of a microscopic theory of
Ezawa et.al.,
in which the Coulomb interaction term appears expilicitly \cite{Ezaw92}.
Later we will take  the particle limit to obtain
the many-body quantum  mechanical wave function.

We    are  interested  in   the  ground   state   wave fucntion
describing FQHE at the filling factor $\nu =1/n$,
therefore $\rho=eB/2   \alpha$.
It has also been   known
that a kind   of Meissner  effect   occurs in the fractional
quantum Hall liquid    system
\cite{Zhan92},  which,  in this formulation,  can   be stated as
 $a_i  -eA_i    = \delta a_i =0$.
 It   is easy   to  show  that above two conditions are
unique constant solution satisfying the equations  of
motion  coming      from
  the   Hamiltonians with as well as without the Coulomb interaction
term \cite{Ezaw92}.

We now take an ansatz for the boson field operator that is known to be good
for the strongly correlated cases \cite{Saki85}
\begin{equation}
\psi = {\rm e}^{i \theta (x)} \sqrt{\rho + \eta (x)},
\end{equation}
where   $\theta   (x)$  is   the phase   operator   and   $\eta  (x)$   is
the  density fluctuation  operator.
Solving the Gauss constraint equation, we get
\begin{equation}
  \delta    a_i  (x)   = -     \frac{\alpha}{\pi} \epsilon_{ij}
\int d^2 y \frac{x^j  - y^j}{(x-y)^2} \eta (y).
\end{equation}
 Note     that   the     statistics  parameter  $\alpha$
  determines    the  strength     of    statistical
 interaction ($\delta a_i (x) \, \delta a_i (y) \sim \alpha^2
\eta (x) \ln |x-y|  \eta (y)$) , and  therefore
considering     fermion   case
 means that the interaction is strong \cite{disc1}. This justifies our
choice of ansatz.

We substitute   the  ansatz    in  the Hamiltonian.
Since   fluctuations are  very
small  in the liquid we  keep     terms only  up    to quadratic
order   of $\eta(x)$  and  $\theta (x)$. This is just the static
version of harmonic
approximation used in Ref. \cite{Zhan92}.
 Then the resulting quadratic Hamiltonian is
\begin{eqnarray}
H  = \int   d^2   x      \Big[  &   &   \frac{\rho}{2m} \partial_j
\theta (x) \partial_j \theta (x) + \frac{1}{8m  \rho}
\partial_j    \eta  (x)  \partial_j  \eta   (x)  \nonumber \\
   &-& \frac{\rho   \alpha}{m    \pi}  \int   d^2 y    \eta   (x)  \ln
|x-y|  \eta (y) + \frac{\alpha}{m} \eta (x) \eta (y) \nonumber \\
   &+&      \frac{e^2}{2    \epsilon}      \int  d^2      y       \eta
(x)   \frac{1}{|x-y|} \eta (y) \Big] .
\end{eqnarray}
In the quadratic Hamiltonian the first three, the fourth and the last
terms come from the kinetic term, the hard-core interaction term and
the    Coulomb interaction  term, respectively.

We next Fourier transform the quadratic Hamiltonian. The transfomation
rule is
\begin{eqnarray}
f(k) &=& \frac{1}{2 \pi} \int d^2 x {\rm e}^{ikx} f(x), \nonumber \\
f(x) &=& \sum^{k_{max}}_{{\bf k} \neq 0} \frac{1}{2 \pi} {\rm e}^{-ikx} f(k).
\end{eqnarray}
We here introduce the   cut-off   of   momentum $k_{max}
(= \sqrt{8 \pi \rho} )$    to
  preserve   the number   of degrees of freedom given by the particle number
 in  the      original  many    body    problem.
We also  exclude    ${\bf k}=0$   case   in  the  summation, since
this  mode   can    be
absorbed in the   condensate, the constant part of
$\psi^\dagger\psi$.
The Fourier  transformed Hamiltonian is
then given by
\begin{eqnarray}
H  =  &   &   \sum^{k_{max}}_{{\bf k} \neq  0}  \Big[   \frac{\rho}{2m}
 k^2   \theta        (-k)  \theta (k) \nonumber \\
 &+& \left(\frac{1}{8m    \rho} k^2+ \frac{\alpha}{m}
+   \frac{\pi    e^2}{\epsilon} \frac{1}{k}     +
\frac{2  \rho  \alpha^2}{m}       \frac{1}{k^2}  \right)
\eta (-k) \eta (k) \Big] .
\end{eqnarray}
We note that a collective field theory, in which Fourier transforms of
the density operator are regarded as collective fields, gives the same
quadratic Hamiltonian in the large-$N$ ($N$ is the number of particles)
expansion \cite{Oka93}.

     In the    non-relativistic   quantum
field   theory    the   field  operators    $\psi  (x)$  and
$\psi^{\dagger}    (x)$    satisfy  the   following  commutation
 relations:
\begin{eqnarray}
{}~& &[    \psi     (x), \psi  (y)]    = [\psi^{\dagger} (x), \psi^{\dagger}
(y)]  =0, \nonumber \\
 & & [ \psi (x),  \psi^{\dagger} (y)] = \delta (x-y).
\end{eqnarray}
We  remember   that    $\psi    (x)$     and  $\psi^{\dagger}  (x)$
have  been  traded by $\eta(x)$     and   $\theta (x)$
through  the  ansatz.     Since
we  consider  here   the case    where   $\eta  (x)$   is   much
smaller    than
$\rho$  we  may    get  an approximate representation
$\psi   (x)     \simeq
  {\rm    e}^{i  \theta      (x)}  (1+    \eta      /2   )
  \sqrt{\rho}$. From these,
 the  following  commutation    relations   of  $\eta (x)$ and $\theta
(x)$  are obtained
\begin{eqnarray}
{}~& &[ \theta  (x), \theta  (y)] =  [  \eta (x), \eta (y)] =0,  \nonumber \\
 & &[ \theta (x), \eta (y)]  = -i \delta (x-y).
\end{eqnarray}
{}From    these   commutation relations  we    may   write    down    directly
commutation relations of $\eta (k)$ and $\theta (k)$
\begin{eqnarray}
{}~& &[ \theta   (k), \theta  (k^{'} )]   =[ \eta   (k), \eta  (k^{'} )]
=0, \nonumber \\
 & & [ \eta (k), \theta (k^{'})] =i \delta_{k+k^{'}} .
\end{eqnarray}
Due  to   these    commutation   relations
the     Fourier   transformed   Hamiltonian  can  be  interpreted
as  the   Hamiltonian
of  a   system  of  $2N$     uncoupled  oscillators  with
the    following   mass  $M_k$   and  natural
    frequency  $\omega_k$  for each mode
\begin{eqnarray}
M_k &=& m/( \rho k^2 ), \nonumber \\
\omega_k &=& \sqrt{\omega^{2}_{c} +2 \pi \frac{\rho}{m}
\frac{e^2}{\epsilon} k
+ \frac{2 \rho \alpha}{m^2} k^2 + \frac{1}{4m^2} k^4} ,
\end{eqnarray}
where $\omega_c (=eB/m)$ is the cyclotron frequency.

Recall that    a   simple  harmonic oscillator  with   Hamiltonian
$h=p^2 /2M    +  (1/2)   M    \omega^2 x^2$       and
  commutation        relations
$[x,x] =0=    [p,p],   ~    [x,p]=i$,  has  the       GSWF
$\Psi_0   \propto \exp(-1/2 ~    M  \omega   x^2      )$.
Using this  harmonic oscillator analogy \cite{Zhan92} we reach at
the GSWF $\Psi_C$ of our system
\widetext
\begin{eqnarray}
\Psi_C  & \propto
&  \exp \left[  - \sum^{k_{max}}_{{\bf k}  \neq   0}  \frac{1}{2} M_k
\omega_k \eta (k) \eta (-k) \right] \nonumber \\
       &     \simeq
&       \exp    \left[     -     \sum^{k_{max}}_{{\bf k}     \neq
    0}       \left( \alpha    \frac{1}{k^2}  +   \frac{\pi
\lambda l_B}{2}   \frac{1}{k}        +  \left(  \frac{\alpha l^{2}_{B}}{2}
- \frac{\pi^2}{8 \alpha} l^{2}_{B} \right) \right) \eta (k) \eta (-k)
\right],
\end{eqnarray}
\narrowtext
\noindent where $l_B ( \equiv 1/ \sqrt{eB} )$ is the
magnetic  length and  $\lambda (\equiv  (e^2/  \epsilon   l_B   )/
\omega_c        )$     is    the  so-called   Landau-level      mixing
parameter measuring  the   effective strength  of the   Coulomb interaction.
It   is very  easy           to  rewrite $\omega_k$  as
   $\omega_k    =  \omega_c  \sqrt{1+ 2  \nu \sqrt{\nu}    \lambda   p
 +   4    \nu p^2   + 4  \nu^2   p^4}$ with  $p \equiv   k/k_{max}$.
Since    $p      \leq          1$  the
apporximation      $\omega_k   \simeq  \omega_c [1 + \nu \sqrt{\nu}
    \lambda        p    +       O(p^2)]$  converges
as  long     as  $\lambda \leq      (   \nu  \sqrt{\nu}    )^{-1}$.
Even for the   massive hole case,
$\lambda$ is $3 \sim 5$ \cite{Sant92,Ando85} and thus this approximation
is good   even for   $\nu     =1/3$ case.
We emphasize that  $p^M  ~(M     \geq       2)$
terms will give only contact terms of no
 physical importance.

We    now         Fourier    transform  $\Psi_C$  back           to      the
coordinate       space.
\begin{eqnarray}
\Psi_C =  \exp   \Big[   &     & \frac{\alpha}{2 \pi}  \int  d^2    x \,
 d^2  y
\,   \eta  (x) \, \ln |x-y| \, \eta (y) \nonumber \\
 &  -     &    \frac{\lambda    l_B}{4}   \int   d^2    x  \,  d^2  y  \,
\eta   (x)    \, \frac{1}{|x-y|} \, \eta (y) \nonumber \\
& - & \left( \frac{\alpha l^{2}_{B}}{2} - \frac{\pi^2}{8 \alpha} l^{2}_{B}
\right) \int d^2 x \, \eta (x) \, \eta (y)  \Big] .
\end{eqnarray}
At this      stage the    GSWF  is expressed in terms     of fields.
To obtain    the     many-body quantum      mechanical    GSWF,
we   take the   particle limit
\begin{equation}
\eta (x) \to \sum^{N}_{a=1} \delta (x-x_a ) - \rho .
\end{equation}
{}From Eqs. (13) and (14)     we  get
\widetext
\begin{eqnarray}
\Psi_C    \propto   \prod_{a<b} |z_a- z_b|^{1/\nu}
\exp   \left[  -  \frac{1}{4} \sum_a |z_a |^2  -  \frac{\lambda}{2}
   \sum_{a<b}  \frac{1}{|z_a -  z_b  |} \right] ,
\end{eqnarray}
\narrowtext
\noindent where $z_a \equiv    x_a /     l_B$.
We neglected the contact terms ($\sim \delta (z_a - z_b )$ and its
derivatives) as well as the
terms coming from the finite size effect, which break the
translational symmetry of the original Hamiltonian.

The $\Psi_C$ is  the wave function of  charged boson.
After (singular) gauge transformation
to get the  wave function of the original spin polarized fermions
the final  form  can be  written  as
\begin{equation}
\Psi_C   = \Psi_{\rm  L}   \times         \exp
     \left[-\frac{\lambda}{2}
 \sum_{a<b} \frac{1}{|z_a  - z_b |} \right],
\end{equation}
where $\Psi_{\rm L}$      denotes the Laughlin wave function.
This   is   the main result of our work.

We now calculate energies for $\Psi_L$ and $\Psi_C$ for $N=5$. If we
do not consider the particle-background and background-background
Coulomb interactions, the Hamiltonian is
\begin{equation}
H = \frac{N}{2} + 2 \sum^{N}_{a} \phi^{\dagger}_a \phi_a + \lambda
\sum_{a<b} \frac{1}{|z_{ab} |} ,
\end{equation}
where $\phi = \partial_z + \bar{z} /4$ and $\phi^{\dagger} =
- \partial_{\bar{z}} + z/4$. Here we take $\omega_c =1$. $(H-N/2)
\Psi_C = H_C \Psi_C$ can be obtained analytically
\begin{eqnarray}
H_C =  & & \frac{5}{4} \lambda \sum^{N}_{a<b} \frac{1}{|z_{ab}|}
{}~-~ \sum^{N}_{a<b} \Big( n- \frac{1}{2} \Big)
\frac{\lambda}{|z_{ab}|^3}
{}~-~ \frac{\lambda^2}{4} \sum^{N}_{a<b}
\frac{1}{|z_{ab}|^4} \nonumber \\
      &-& \frac{\lambda^2}{8} \sum^{N}_{a \neq b \neq c}
\frac{{\mbox Re} (z_{ab} \bar{z}_{ac} )}{|z_{ab} z_{ac}|^3}
{}~-~ \frac{n \lambda}{2} \sum^{N}_{a} \sum^{N}_{b \neq a}
\sum^{N}_{c \neq a,b} \frac{z_{ab}}{|z_{ab}|^3}
\frac{1}{z_{ac}} .
\end{eqnarray}
Energies for $\Psi_L$ and $\Psi_C$, which are defined by $\int d^2 z
\Psi_{L}^{*} \Psi_L ( \sum_{a<b} \lambda /|z_{ab}|)/ \int d^2 z
\Psi_{L}^{*} \Psi_{L}$ and $ \int d^2 z \Psi_{C}^{*} \Psi_C H_C /
 \int d^2 z \Psi_{C}^{*} \Psi_C$, are calculated for $N=5$,
$\lambda =2/3$ case using Monte Carlo
method, and are given by 2.05 and 1.75, respectively. That is,
the energy is lowered by about $15 \%$. This  positive result from
the five-body calculation provides a motivation for more
numerical efforts. There is another support for the would-be
superiority of our wave function.   The total energy
per particle is given in terms of the radial distribution function $g(|z|)$.
Laughlin argued that the main reason why the fractional quantum Hall liquid
is variationally superior to the Wigner crystal at $\nu = 1/3$ is
the fact that
$g(|z|)$ for the liquid is pushed out further from the origin
than $g(|z|)$ for the crystal \cite{Laug90}.
It is evident that the modifying term in our trial wave
function further push out $g(|z|)$ from the origin.
Thus we expect our wave
function will be variationally superior to the Laughlin wave function.

Based on our improved ground state wave function,
 we can directly write down the wave functions
$\Psi_C \,^{-z_0}$ and $\Psi_C \,^{+z_0}$ for the Laughlin quasihole and
quasiparticle excitations following the Laughlin's
argument \cite{Laug83,Laug90,Halp83} of adiabatic flux insertion
\begin{eqnarray}
\Psi_C \,^{-z_0} &=& \exp \Big[ - \frac{1}{4} \sum_a |z_a |^2 -
\frac{\lambda}{2} \sum_{a<b} \frac{1}{|z_a - z_b |} \Big]
 \prod_a (z_a - z_0 ) \prod_{a<b} (z_a - z_b )^m , \nonumber \\
\Psi_C \,^{+z_0} &=& \exp \Big[ - \frac{1}{4} \sum_a |z_a |^2 -
\frac{\lambda}{2} \sum_{a<b} \frac{1}{|z_a - z_b |} \Big]
\prod_a \Big( 2 \frac{\partial}{\partial z_a} - z_{0}^{*} \Big)
\prod_{a<b} (z_a - z_b )^m .
\end{eqnarray}
One might try to derive these wave functions using the mean field method
used above from a vortex field theory obtained via the
Hubbard-Stratonovich transformation \cite{Ezaw92,Zhan92}.
However this is not so promising,
since the linear approximation (or the harmonic approximation) is
not good for the intrinsically nonlinear object like a vortex and moreover
the vortex size is comparable with the magnetic length, which is
a typical length of the system.

The appearance of the FQHE at $\nu=1/7$ has been controversal.  Goldman et.
al's
experiment \cite{Lam84} and Jain's classification \cite{Jain89} insist on
the appearance. However, Lam and Girvin's classic work \cite{Lam84} and the
most updated numerical calculation \cite{Pric93,Zhu93} show that the
critical filling factor $\nu_c$ for the liquid-solid transition is larger
than $1/7$. In Ref.\cite{Pric93}, they used a trial wave function
including the Landau level mixing effect, which however has a form different
from ours in (16).
It is interesting to see if our improved wavefunction
gives the ground state
energy lower than that of Ref.\cite{Pric93} making $\nu_c$ smaller.
In case $\nu_c$ is less than $1/7$ with our modified wave function,
then the appearance of the FQHE
at $\nu=1/7$ could be explained theoretically.

As discussed in the beginning of this paper, the Laughlin's wave
functions for the ground state, quasihole and quasiparticle excitations
are independent of $\lambda$ and therefore cannot explain the
experimentally observed reduction of energy gap between the ground state
and the Laughlin quasiexciton as $\lambda$ increases. However, our wave
functions in (16) and (19) are explicitly dependent on $\lambda$ and
therefore there is a chance to explain the reduction of the energy gap.

In conclusion we propose systematically derived improved
trial wave functions for the ground state, and
Laughlin quasihole and quasiparticle excitations. These include the
Landau level mixing effect explicitly.
Since the validity of our wavefunction will shed new light on the
appearance of the fractional quantum
Hall effect at $\nu = 1/7$ \cite{Lam84}
and the reduction of energy gap as the  effective strength
of Coulomb interaction increases \cite{Sant92,Yosh84},
it will be very interesting to continue the numerical study.

This work was supported in part by the KOSEF and
by the Korean Ministry of Education.


\begin{references}
\bibitem{disc1}For a near-boson case ($\alpha \simeq 0$) the ansatz
$\psi = \sqrt{\rho} + \eta$ ($\eta$ is complex) is more appropriate. Using
this ansatz and keeping terms up to quadratic order of $\eta (x)$ we may
get the GSWF (the Jastrow wave function)
for charaged superfluid, $\Psi^{\nu}_{s} = \prod_{a<b}
(z_a - z_b )^{1/ \nu}$ (see Ref. 11).
\end{references}
\end{document}